\documentstyle[preprint,aps]{revtex}

\begin{document}

\draft
\title{Quantum Limits for Measurements on Macroscopic  \\
Bodies : a Decoherence Analysis}
\author{Carlos O. Escobar, \ L\'ea Ferreira dos Santos
\\and \\Paulo C. Marques\ \thanks{e-mail : pacmarques@uspif.if.usp.br}}
\address{Instituto de F\'\i sica da Universidade de S\~ao Paulo\\
01452-990 C.P. 20516--S\~ao Paulo, SP\\
Brazil\\}
\maketitle

\begin{abstract}
We consider in this paper the quantum limits for measurements on macroscopic
bodies which are obtained in a novel way employing the concept of decoherence
coming from an analysis of the quantum mechanics of dissipative systems.
Two cases are analysed, the free mass and the harmonic oscillator, and
for both systems we compare our approach  with previous treatments of such
limits.
\end{abstract}

\section{Introduction}

One of the outstanding problems in contemporary physics is the detection
of the gravitational waves predicted by Einstein in his General Relativity
Theory~\cite{gr}. The detection of such waves presents a challenge
to the experimentalists who have to monitor
extremely small fluctuations in the
position of macroscopic bodies~\cite{rev}. The sensitivity planned for the
future generation of detectors, will bring us close to the limits imposed
by quantum mechanics~\cite{vb}.

It is our purpose in this paper to examine in detail the determination of
quantum limits for free masses and for the harmonic oscillator, subjected
to thermal fluctuations. We rely on the methods developed by several authors
to analyse the quantum mechanics of a system in interaction with its
environment~\cite{cl}.

We differ in this respect from Braginsky~\cite{vb} who
uses a two step procedure in order to obtain the quantum limits, firstly
considering the limits imposed by the uncertainty principle on an isolated
system, not subjected to thermal fluctuations and then in the second step
looking at the same system, this time behaving classically but interacting with
the environment.

This procedure neglects the fact that the system is always in
contact with the environment. We remove this asymmetry treating the system
quantum mechanically and subjected to thermal fluctuations from the very
beginning.

We want to investigate the conditions under which the quantum behavior of a
macroscopic body cannot be ignored or equivalently for our problem,when
classical behavior sets in.This problem has become of importance in
recent studies of quantum cosmology~\cite{Hu}
as well as in the quantum theory of the
measurement process~\cite{Zurek}. The concept of decoherence arising
from these studies will be
central in our considerations. The way we see the onset of quantum behavior of
a macroscopic body is related to the decoherence time scale in the following
way : {\it A macroscopic test body must be treated quantum mechanically
if the time
interval during which a measurement of some of its observables is made, is
smaller than the decoherence time}.

In section II we briefly review the concept of decoherence as used  by
Zurek and collaborators~\cite{Zu}. In section III we use the decoherence
time scale in
order to obtain the conditions for quantum behavior of a macroscopic body in
two cases : the harmonic oscillator and the free mass (free in the sense
it is not subjected to an external field but, of course, in
contact with a reservoir). We end section III with a criticism of
Braginsky's derivation of the quantum condition
for a free mass. In section IV we conclude the paper outlining some future and
open problems.

\section{Decoherence Time Scale}

When a system interacts with a thermal reservoir, it is usual to analyse its
evolution in terms of the reduced density
matrix ($\rho_r$), obtained by coarse
graining from the full density matrix ($\rho$) for the system plus reservoir.
This coarse graining is effected by tracing out the degrees of freedom of the
environment :

\begin{equation}
\rho_r = {\rm Tr}_E [ \rho ]
\end{equation}

where {\small $E$} refers to the environment variables.

Modelling the reservoir as an ensemble of harmonic oscillators interacting
weakly with the system, it is possible to approximate the
system-reservoir coupling by a linear interaction. This system
can be studied by functional methods following the pioneering approach of
Feynman and Vernon~\cite{Feyver} who showed that the effect of the reservoir
on the time evolution of the system is summarized in a
functional, the so-called influence functional~\cite{Feyhib}.

Caldeira and Leggett~\cite{cl} investigated in detail the functional approach
to the quantum Brownian motion showing that the reduced density
matrix $\rho_r$ obeys,
in the high-temperature limit, for an ohmic environment, a master equation
which is of the following form :

\begin{equation}
\frac{d\rho_r}{dt}
= - \frac{i}{\hbar}[H_R , \rho_r]
- \frac{i \gamma}{2 \hbar}[ \{ p , x \} , \rho_r ]
- \frac{2 m \gamma k_B T}{{\hbar}^2}[x , [x , \rho_r]]
- \frac{i \gamma}{\hbar}([x , \rho_r p]
- [p , \rho_r x])
\end{equation}

In the last equation $H_R$ is the renormalized self-Hamiltonian of the system,
${\gamma}^{-1}$ the relaxation time,
$T$ the temperature.

Zurek et al.~\cite{Zu} introduced the linear entropy

\begin{equation}
\varsigma = {\rm Tr} (\rho_r - {\rho_r}^2),
\end{equation}

in order to make quantitative the notion of the decoherence as measured by the
vanishing of off-diagonal terms in $\rho_r$,
which are responsible
for the interference effects.

Using the master equation (Eq.(2))
for an initially pure state that
remains approximately pure during the
course of its evolution - a good approximation for a system with small
dissipation - it is easy to show that :

\begin{eqnarray*}
\dot{\varsigma} & = & 4 D {\rm Tr} ({{\rho_r}^2}x^2
- {\rho_r}x {\rho_r}x )
- 2 \gamma {\rm Tr} {\rho_r}^2 \\
& \simeq & 4 D ( \langle x^2 \rangle - \langle x \rangle^2 ) = 4 D (\Delta x)^2
\end{eqnarray*}

where $D \equiv 2 m \gamma k_B T/ \hbar^2$, and $(\Delta x)^2$ is the mean
square deviation of the position of the system.

This defines a decoherence time scale~\cite{F1}, given by :

\begin{equation}
\tau_D = \frac{\hbar^2}{2 m \gamma k_B T (\Delta x)^2}
\end{equation}

which provides a measure for the amount of time required for the quantum
system to start achieving classical features.

In the following section we establish our criterion for obtaining the
conditions under which a macroscopic body will behave quantum mechanically.

\section{The Limits}

Let us now apply the above concepts in order to obtain the quantum limits for
two systems, the free mass, relevant to laser
interferometric antennas, and the harmonic oscillator, relevant to
mechanical bars.

\subsection{Free Mass}

The quantum limit is obtained when $ (\Delta x)^2 $ in Eq.(4) becomes of the
same magnitude as the uncertainty arising from Heisenberg's principle, when
applied to successive measurement of the position of the free mass
separated by a time interval $ \tau $ . This quantum mechanical uncertainty is
given by~\cite{Caves}

\begin{equation}
(\Delta x)^2 \approx \frac{\hbar \tau}{m}
\end{equation}

Inserting (5) into the expression for $ \tau_D $ ( Eq.(4) ), we obtain

\begin{equation}
\tau_D = \frac{\hbar}{2 \gamma k_B T \tau}
\end{equation}

{\it If} \ \ $ \tau_D $ \ {\it is greater than the time interval between
two successive
measurements}, $ \tau $ , {\it the system must be treated quantum
mechanically}. When
this happens we obtain the quantum limit

\begin{equation}
\hbar > 2 \gamma k_B T \tau^2
\end{equation}

Since $ \gamma^{-1} $ is the relaxation time
of the system ($ \tau^* $), Eq.(7)
can be rewritten in the more usual form

\begin{equation}
\hbar > 2 k_B T \frac{\tau^2}{\tau^*}
\end{equation}

which is the limit obtained by Braginsky~\cite{vb}, who arrived at
this result by a somewhat obscure path, as we will make explicit later on.

\subsection{Harmonic Oscillator}

For a harmonic oscillator of fundamental frequency $ \omega $, the Heisenberg
uncertainty principle gives~\cite{Caves}

\begin{equation}
(\Delta x)^2 \approx \frac{\hbar}{2 m \omega}
\end{equation}

for a measurement time of the order of the period of the harmonic
oscillator $ \left( \tau \approx \frac{2 \pi}{\omega} \right)$.

Replacing (9) into (4) leads to

\begin{equation}
\tau_D = \frac{\hbar \omega}{\gamma k_B T}
\end{equation}

Imposing $ \tau_D > \tau $ gives the quantum limit for the oscillator

\begin{equation}
\hbar > \gamma \frac{k_B T \tau}{\omega} =
\frac{k_B T}{\omega}\frac{\tau}{\tau^*}
\end{equation}

A similar result was obtained by Braginsky~\cite{vb}, only differing from
(11) by a numerical factor.

Having shown how to obtain the quantum limits in an internally consistent way,
which takes into account both quantum and thermal fluctuations, we now raise
one further objection to the derivation by Braginsky of the quantum limit for a
free mass.

The authors of reference~\cite{vb2} use as the starting point Nyquist's
theorem
which gives for the spectral density~\cite{F2} of
the fluctuating force the result :

\begin{equation}
\left< F^2_{fl} \right>_\omega = 4 k_B T \gamma
\end{equation}

Then they use this fluctuating force properly integrated over a range of
frequencies $ \Delta \omega \approx \tau^{-1} $ to find the displacement of
the particle $ x = x_0 + \frac{1}{2}a t^2 $ under a constant
acceleration given by

\begin{equation}
a = \frac{\sqrt{\left< F^2_{fl} \right>_\omega \tau^{-1}}}{m}
\end{equation}

which then gives for the {\it displacement}

\begin{equation}
\Delta x = \sqrt{\frac{k_B T}{m} \frac{\tau^3}{\tau^*}}
\end{equation}

This result is in disagreement with a standard treatment
of the classical Brownian particle~\cite{Reif}. Of course the latter treatment
refers to the mean square deviation of the position of the particle, which as a
matter of fact is what is monitored in gravitational wave antennas~\cite{rev},
while Braginsky considers a displacement to be later on compared with a
fluctuation (from Heisenberg's principle).

Finally we remark that when Braginsky
considers the harmonic oscillator, the classical part is treated in the right
way, as a fluctuating Brownian particle in the potential of a harmonic
oscillator.

\section{Conclusions}

We have obtained the conditions for the quantum behavior of a macroscopic body
using a decoherence analysis, which treats the systems quantum mechanically
from the very beginning, thus removing some inconsistencies present in previous
treatments of the subject.

Zurek et al. showed~\cite{D47} through a numerical study of the time evolution
of the Wigner function that
the decoherence time
scale defined by Eq.(4) is still valid even at low temperatures.

It is important to stress that we do not propose the master equation (Eq.(2))
for describing the time evolution of a mechanical gravitational wave antenna
for two reasons; firstly as it operates at small
temperatures~\cite{rev} and secondly because the dissipation
mechanism for such
antennas is fairly complicated~\cite{vb2} hardly being ohmic.
 We have considered
in this paper a simplified model for a Weber bar and exploited some
of the  consequences of Eq.(2) for the quantum limits of such a system.

We are currently working on an analysis of quantum non-demolition
measurements~\cite{Ho}, taking into account dissipative effects.

\acknowledgments
L.F.dos Santos and P.Marques F. would like to thank Funda\c{c}\~{a}o
de Amparo \`a Pesquisa do Estado de S\~{a}o Paulo for financial support.  C.O.
Escobar wishes to thank  Conselho Nacional de Desenvolvimento
Cient\'{\i}fico e Tecnol\'ogico for partial financial support.


\begin{references}
\bibitem[*]{F1} As a matter of fact the decoherence time depends on the initial
                state of the system. However we always choose the smallest
                possible time scale for decoherence given by Eq.(4), as
                emphasized by Zurek et al.~\cite{D47}. In so doing we are just
                relaxing the time condition of our decoherence criterion.
\bibitem[**]{F2} The use of this spectral density entails a restricted limit of
                 applicability, namely a high-temperature limit, where quantum
                 fluctuations are completely neglected. It is unsuitable
                 for studying the transition from classical to quantum
                 behaviour to use it as a starting point.
\bibitem{gr} A. Einstein, Sitzungsber Preuss.Akad.Wiss., 688 (1916).
\bibitem{rev} For a review of the current status of G.W. detectors see,
              {\it The detection of Gravitational Waves} ed. D.G. Blair,
              Cambridge University Press, 1991.
\bibitem{vb} V.B. Braginsky and Yu.I. Vorontsov, Sov. Phys Usp. {\bf 17},
             644 (1974); V.B.Braginsky, Sov. Phys. Usp. {\bf 31}, 836 (1988).
\bibitem{cl} A.O. Caldeira and A.J. Leggett, Physica A {\bf 121}, 587 (1983).
             W.G. Unruh and W.H. Zurek, Phys.Rev. D {\bf 40}, 1071 (1989).
             W.H. Zurek. Phys.Today {\bf 44},No 10, 36 (1991)
             (for an introduction).
\bibitem{Hu} B.L. Hu, J. P. Paz and Y. Zhang, Phys. Rev. {\bf D45},
             2843 (1992); {\bf D47}, 1576 (1993). M. Gell-Mann and J.B.
             Hartle in~\cite{Zurek}.
\bibitem{Zurek} {\it Complexity, Entropy and the Physics of Information}, ed.
                W. Zurek, Vol IX (Addison-Wesley, Reading, 1990).
\bibitem{Zu} W.H. Zurek, S. Habib and J.P. Paz, Phys.Rev.Lett., {\bf 70}, 1187
             (1993).
\bibitem{Feyver} R.P. Feynman and F.L. Vernon, Ann. Phys. {\bf 24}, 118 (1963).
\bibitem{Feyhib} R.P Feynman and A.R. Hibbs, {\it Quantum Mechanics and Path
                 Integrals}, McGraw-Hill Book Company, 1965.
\bibitem{D47} J.P.Paz, S. Habib and W.H. Zurek, Phys. Rev. {\bf D47},
              488 (1993).
\bibitem{Caves} C.M. Caves, K.S. Thorne, R.W.P. Drever, V.D. Sandberg,
                M. Zimmermann. Rev. Mod. Phys. {\bf 52}, 341 (1980).
\bibitem{vb2} V.B. Braginsky and A.B. Manukin, {\it Measurements of Weak
              Forces in
              Physical Experiments} , Univ. of Chicago Press, 1977.
              V.B. Braginsky, V.P. Mitrofanov, V.I.Panov, {\it Systems with
              Small Dissipation}, Univ. of Chicago Press, 1985.
\bibitem{Reif} F. Reif, {\it Fundamentals of Statistical and Thermal Physics},
               McGraw-Hill, 1965.
\bibitem{Ho} J.N. Hollenhorst, Phys.Rev. D {\bf 19},1669 (1979).
\end{references}
\end{document}